\documentclass{article}
\usepackage{booktabs}
\usepackage{amsmath,graphicx,mlspconf}
\usepackage{amssymb} 
\usepackage{multirow}
\usepackage{algorithm}
\usepackage{algpseudocode}
\usepackage{caption}
\title{Whisper Speaker Identification: Leveraging Pre-Trained Multilingual Transformers for Robust Speaker Embeddings}

\name{Jakaria Islam Emon$^{1}$, Md Abu Salek$^{1}$, Kazi Tamanna Alam$^{2}$}
\address{$^{1}$ Hokkaido Denshikiki Co., Ltd., Sapporo, Japan \\
         $^{2}$ Barisal Information Technology College (BITC), Barisal, Bangladesh}
\begin{document}

\maketitle

\begin{abstract}
Speaker identification in multilingual settings presents unique challenges, particularly when conventional models are predominantly trained on English data. In this paper, we propose \textit{WSI} (Whisper Speaker Identification), a framework that repurposes the encoder of the \textit{Whisper} automatic speech recognition model pre-trained on extensive multilingual data to generate robust speaker embeddings via a joint loss optimization strategy that leverages online hard triplet mining and self-supervised Normalized Temperature-scaled Cross Entropy (nt-xent) loss. By capitalizing on Whisper’s language-agnostic acoustic representations, our approach effectively distinguishes speakers across diverse languages and recording conditions. Extensive evaluations on multiple corpora, including VoxTube (multilingual), JVS (Japanese), CallHome (German, Spanish, Chinese, and Japanese), and Voxconverse (English), demonstrate that WSI consistently outperforms state-of-the-art baselines, namely Pyannote Embedding, ECAPA-TDNN, and X-vector, in terms of lower equal error rates and higher AUC scores. These results validate our hypothesis that a multilingual pre-trained ASR encoder, combined with joint loss optimization, substantially improves speaker identification performance in non-English languages.

\end{abstract}

\begin{keywords}
Speaker Identification, Self-Supervised Loss, Whisper, Open-Set Speaker Recognition, Speaker Identification.
\end{keywords}

\section{Introduction}
Recent advances in automatic speech recognition (ASR) have been driven by large-scale, pre-trained transformer-based models such as Whisper \cite{radford_robust_2022}. These models achieve state-of-the-art performance in multilingual transcription by generating robust and generalized representations of spoken language. For example, Peng et al. \cite{peng_improving_2023} demonstrated that such models excel not only in transcribing diverse speech content but also in capturing intricate linguistic nuances, thereby broadening the scope of ASR applications.

Despite these successes, the potential of pre-trained ASR models for speaker-centric applications such as speaker identification, verification, and diarization remains under-explored. Traditional speaker recognition methods typically rely on speaker embeddings generated by deep neural networks trained on extensive datasets. Examples include x-vector embeddings \cite{fathan_impact_2024} and Lepage et al. \cite{lepage_experimenting_2023}, Wespeaker-voxceleb-resnet34-LM introduced by Wang et al. \cite{wang_wespeaker_2023}, ECAPA-TDNN proposed by Desplanques et al. \cite{desplanques_ecapa-tdnn_2020}, and Pyannote embeddings by Bredin et al. \cite{bredin_pyannoteaudio_2023}. In addition, Nagrani et al. \cite{nagrani_voxceleb_2020} leveraged large-scale datasets like VoxCeleb to train models capable of distinguishing between known and unseen speakers. However, these approaches often experience performance degradation in multilingual environments where speakers may switch between languages or dialects that are underrepresented in the training data. 
A key challenge in multilingual speaker recognition is ensuring that speaker embeddings remain language-agnostic \cite{kim_speaker-attributed_2024,xuan_multi-scene_2022}. Chen et al. \cite{chen_large-scale_2022} showed that embeddings capturing speaker characteristics independent of the spoken language enable consistent identification across diverse linguistic contexts. Although Lepage et al. \cite{lepage_experimenting_2023} suggest that deep neural network-based embeddings can inherently capture language-independent speaker traits, Song et al. \cite{song_introducing_2024} report that language-specific features may inadvertently affect these embeddings, thereby compromising their robustness. Moreover, the adaptation of pre-trained ASR models for speaker recognition tasks has shown promise \cite{peng_fine-tune_2024}. Kanda et al. \cite{kanda_end--end_2021} and Sang and Hansen \cite{sang_efficient_2024} demonstrated that transformer-based architectures can effectively capture language-agnostic acoustic features, leading to improved speaker discriminability in multilingual contexts.

In this work, we build upon recent advances to improve speaker recognition in linguistically diverse scenarios. We propose WSI, a framework that repurposes pre-trained transformer-based speech embeddings for generating discriminative speaker representations via a joint loss optimization strategy that leverages online triplet mining and self-supervised NT-Xent\cite{chen_simple_2020} losses.
Our main contributions can be summarized as follows: 
\begin{itemize}

    \item \textbf{Repurposing Pre-Trained Transformer-Based ASR Models for Speaker Embeddings:} We leverage a pre-trained Whisper encoder to extract robust acoustic representations and repurpose them for speaker verification. The encoder is fine-tuned jointly with a projection head using a combined loss objective. This approach effectively utilizes existing acoustic knowledge, eliminating the need to train a speaker model from scratch (see Algorithm~\ref{alg:training}).
    
    \item \textbf{Joint Loss Optimization for Enhanced Speaker Discrimination:} Our method jointly optimizes an online hard triplet loss and a self-supervised NT-Xent loss to learn robust and discriminative speaker embeddings. For instance, on the VoxTube dataset, our approach achieves an Equal Error Rate (EER) of 0.90\%, which is substantially lower than those of competing methods (Pyannote: 3.38\%, ECAPA-TDNN: 1.17\%, and X-vector: 7.23\%), (see Figure~\ref{fig:eer_boxplot}).

    \item \textbf{Multilingual Open-Set Speaker Identification:} Unlike conventional models that may underperform in multilingual settings, our framework inherently supports open-set scenarios across multiple languages. For example, on the CallHome corpus, WSI achieves EERs of 10.50\% in German, 11.20\% in Spanish, 12.00\% in Chinese, and 10.80\% in Japanese, substantially outperforming competing methods. These results confirm the robustness and generalizability of our approach across diverse linguistic contexts (see Table \ref{tab:combined_results}).

\end{itemize}

The remainder of this paper is organized as follows. Section~II describes the proposed methodology. Section~III details the experimental setup, including dataset descriptions and evaluation metrics. Section~IV presents the results and discussion, and Section~V concludes the paper with future work directions.

\section{Method}
In this paper, we propose a discriminative speaker embedding framework for open-set speaker verification that leverages a pre-trained Whisper encoder as a robust embedding extractor. The encoder is fine-tuned jointly with a projection head using a combined loss objective that integrates an online hard triplet loss with a self-supervised NT-Xent loss. The additional self-supervised loss enforces consistency across different augmented views, thereby enhancing the robustness of the learned embeddings.

\subsection{Network Architecture}
\begin{figure*}[t]
    \centering
    \includegraphics[width=\linewidth]{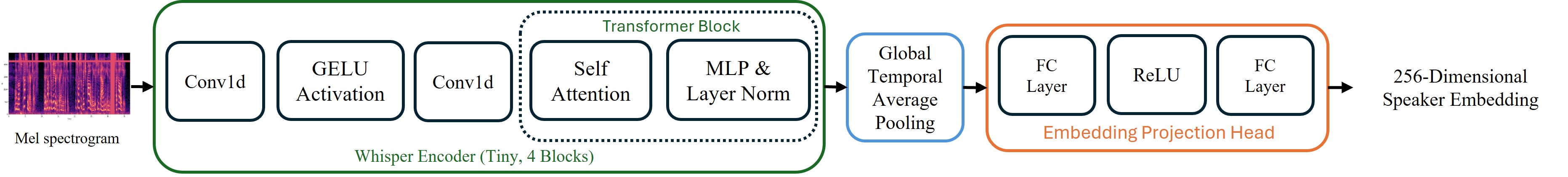}
    \caption{WSI Architecture}
    \label{fig:architecture}
\end{figure*}
Our network comprises two main components:
\begin{enumerate}
    \item \textbf{Whisper Encoder:} Given an input log-mel spectrogram $\mathbf{X} \in \mathbb{R}^{F \times T}$, the encoder extracts frame-level embeddings:
    \begin{equation}
    \mathbf{E} = \{\mathbf{e}_1, \mathbf{e}_2, \dots, \mathbf{e}_T\}, \quad \mathbf{e}_t \in \mathbb{R}^{D},
    \label{eq:frame_embeddings}
    \end{equation}
    where $F$ is the number of frequency bins, $T$ is the number of time frames, and $D$ is the output dimensionality of each frame-level embedding. These embeddings are then aggregated via global mean pooling:
    \begin{equation}
    \bar{\mathbf{e}} = \frac{1}{T}\sum_{t=1}^{T}\mathbf{e}_t,
    \label{eq:mean_pooling}
    \end{equation}
    yielding an averaged feature vector of the input audio segment.
    
    \item \textbf{Projection Head:} The pooled representation is transformed via a projection head $f_{\text{proj}}(\cdot)$ into a compact speaker embedding:
    \begin{equation}
    \mathbf{z} = f_{\text{proj}}(\bar{\mathbf{e}}), \quad \mathbf{z} \in \mathbb{R}^{256},
    \label{eq:speaker_embedding}
    \end{equation}
    where $\mathbf{z}$ is the final speaker embedding.
\end{enumerate}
Figure~\ref{fig:architecture} illustrates the detailed network architecture.

\subsection{Joint Loss Optimization }
To learn discriminative and robust embeddings, we employ a joint training objective that combines an online hard triplet loss with a self-supervised NT-Xent loss \cite{chen_simple_2020}. For each input audio sample $x_i$, two augmented views are generated: a noise-augmented version $x_i^{(n)}$ and a time-stretched version $x_i^{(t)}$. The embeddings are computed as follows:
\begin{equation}
\mathbf{z}_i = f_{\text{proj}}\Biggl(\frac{1}{T}\sum_{t=1}^{T} f_{\text{enc}}(\mathbf{X}_{i})\Biggr),
\label{eq:embed_original}
\end{equation}
\begin{equation}
\mathbf{z}_i^{(n)} = f_{\text{proj}}\Biggl(\frac{1}{T}\sum_{t=1}^{T} f_{\text{enc}}(x_i^{(n)})\Biggr),
\label{eq:embed_noise}
\end{equation}
\begin{equation}
\mathbf{z}_i^{(t)} = f_{\text{proj}}\Biggl(\frac{1}{T}\sum_{t=1}^{T} f_{\text{enc}}(x_i^{(t)})\Biggr).
\label{eq:embed_time}
\end{equation}
The online hard triplet loss is defined as:
\begin{equation}
\mathcal{L}_{\text{triplet}} = \max\Bigl(0,\, m + \|\mathbf{z}_A - \mathbf{z}_P\|_2 - \|\mathbf{z}_A - \mathbf{z}_N\|_2\Bigr),
\label{eq:triplet_loss}
\end{equation}
where $\mathbf{z}_A$, $\mathbf{z}_P$, and $\mathbf{z}_N$ denote the embeddings of the anchor, positive, and negative samples, respectively, and $m=1.0$ is the margin.

Along with that, we enforce consistency between the original and augmented views using the NT-Xent loss:
\begin{equation}
\mathcal{L}_{NT} = \frac{1}{2}\Bigl[\text{NTXent}(\{\mathbf{z}_i\}, \{\mathbf{z}_i^{(n)}\}) + \text{NTXent}(\{\mathbf{z}_i\}, \{\mathbf{z}_i^{(t)}\})\Bigr].
\label{eq:nt_xent_loss}
\end{equation}
The overall training objective is a weighted combination of the two losses:
\begin{equation}
\mathcal{L} = \mathcal{L}_{\text{triplet}} + \lambda\, \mathcal{L}_{NT},
\label{eq:total_loss}
\end{equation}
where $\lambda$ is the weight balancing the self-supervised loss.

During training, only the Whisper encoder and the projection head are updated. 
\begin{algorithm}[!ht]
\caption{Training Procedure for Online Triplet Mining with Multi-View Self-Supervision}
\label{alg:training}
\begin{algorithmic}[1]
\State \textbf{Input:} Training dataset $\mathcal{D}=\{(x_i,y_i)\}$, learning rate $\eta$, margin $m$, self-supervised weight $\lambda$, epochs $E$, batch size $B$
\State \textbf{Initialize:} Pretrained Whisper encoder $f_{\text{enc}}$, projection head $f_{\text{proj}}$, Adam optimizer
\For{$epoch = 1$ to $E$}
    \For{each batch $\mathcal{B} = \{(x_i, y_i)\}_{i=1}^{B} \subset \mathcal{D}$}
        \State \textbf{Data Augmentation:}
        \For{each $x_i \in \mathcal{B}$}
            \State $x_i^{(n)} \gets \text{NoiseAugmentation}(x_i)$
            \State $x_i^{(t)} \gets \text{TimeStretch}(x_i)$
        \EndFor
        \State \textbf{Feature Extraction:} 
        \State \quad Original audios: $F = \{f(x_i)\}$
        \State \quad Noise-augmented: $F^{(n)} = \{f(x_i^{(n)})\}$
        \State \quad Time-stretched: $F^{(t)} = \{f(x_i^{(t)})\}$
        \State \textbf{Embedding Computation:}
        \State Compute embeddings for original audios:
        \State \quad $\mathbf{z}_i = f_{\text{proj}}\Big(\text{pool}\big(f_{\text{enc}}(F_i)\big)\Big)$
        \State Compute $\mathbf{z}_i^{(n)}$ and $\mathbf{z}_i^{(t)}$ from $F^{(n)}$ and $F^{(t)}$
        \State \textbf{Loss Computation:}
        \State Compute online hard triplet loss: 
        \[
        \mathcal{L}_{triplet} = \text{TripletLoss}(\{\mathbf{z}_i\}, \{y_i\}, m)
        \]
        \State Compute NT-Xent loss for self-supervision:
        \[
        \mathcal{L}_{NT} = \frac{1}{2}\Big[\text{NTXent}(\{\mathbf{z}_i\}, \{\mathbf{z}_i^{(n)}\}) + \text{NTXent}(\{\mathbf{z}_i\}, \{\mathbf{z}_i^{(t)}\})\Big]
        \]
        \State Compute total loss:
        \[
        \mathcal{L} = \mathcal{L}_{triplet} + \lambda\, \mathcal{L}_{NT}
        \]
        \State Update model parameters using backpropagation with loss $\mathcal{L}$
    \EndFor
\EndFor
\State \textbf{Output:} Trained model parameters
\end{algorithmic}\vspace{-5mm}
\end{algorithm}
Algorithm~\ref{alg:training} summarizes the training procedure.

\subsection{Evaluation}

During inference, each utterance is mapped to a speaker embedding $\mathbf{z}$ using the trained network. The cosine similarity between two embeddings is computed as:
\begin{equation}
\text{sim}(\mathbf{z}_1, \mathbf{z}_2) = \frac{\mathbf{z}_1 \cdot \mathbf{z}_2}{\|\mathbf{z}_1\|\|\mathbf{z}_2\|},
\label{eq:cosine_similarity}
\end{equation}
where $\mathbf{z}_1 \cdot \mathbf{z}_2$ denotes the dot product.

A decision threshold $\tau$ is applied to determine whether two embeddings belong to the same speaker. The system's performance is evaluated using Equal Error Rate (EER) and Area Under the Curve (AUC).

\textbf{Equal Error Rate (EER)} is defined as the operating point where the False Positive Rate (FPR) equals the False Negative Rate (FNR):
\begin{equation}
\begin{split}
\text{EER} &= \text{FPR}(t^*) = \text{FNR}(t^*), \\
t^* &= \arg\min_{t} \left|\text{FPR}(t) - \text{FNR}(t)\right|.
\end{split}
\label{eq:EER}
\end{equation}
The FPR and FNR are computed as:
\begin{equation}
\text{FPR}(t) = \frac{\text{FP}}{\text{FP}+\text{TN}}, \quad
\text{FNR}(t) = \frac{\text{FN}}{\text{FN}+\text{TP}},
\label{eq:FPR_FNR}
\end{equation}
where FP, TN, FN, and TP denote the number of false positives, true negatives, false negatives, and true positives, respectively.

\section{Experimental Setup}
\subsection{Dataset}
Our proposed WSI model was developed and evaluated using multiple speech corpora. For training, we employ the VoxTube dataset \cite{yakovlev_voxtube_2023}, a large-scale corpus derived from YouTube videos suitable for speaker identification task. Each instance in VoxTube is a 4-second audio segment, accompanied by metadata such as a unique speaker identifier (\texttt{spk\_id}) and the primary language (covering 70 languages). Key statistics of the VoxTube dataset are provided in Table~\ref{tab:dataset_stats}. For evaluation, the following additional corpora are used \ref{tab:eval_summary}: 
\begin{itemize}
    \item \textbf{JVS (Japanese Versatile Speech):} A studio-recorded corpus with 100 professional speakers and approximately 30 hours of audio sampled at 24 kHz.
    \item \textbf{CallHome and Voxconverse:} These datasets offer diverse language coverage (German, Spanish, Chinese, Japanese, and English).
\end{itemize}

\begin{table}[h]
\centering
\captionsetup{font=small,justification=raggedright,singlelinecheck=false}
\caption{Key Statistics of the VoxTube Dataset}
\label{tab:dataset_stats}
\small
\begin{tabular}{lrrr} 
\toprule
\textbf{Property}         & \textbf{Train} & \textbf{Validation} & \textbf{Test} \\ 
\midrule
Unique Speakers           & 4,000          & 500                 & 540           \\  
Audio Segments            & 3,500,000      & 450,000             & 489,888       \\  
Total Duration (hours)    & 3,880          & 500                 & 553           \\   
\bottomrule
\end{tabular}
\end{table}
\vspace{-5pt}

\begin{table}[h!]
    \centering
    \captionsetup{font=small,justification=raggedright,singlelinecheck=false}
    \caption{Evaluation Corpus Summary}
    \label{tab:eval_summary}
    \small
    \begin{tabular}{@{}p{2cm}p{3cm}p{2cm}@{}}
    \toprule
    \textbf{Corpus Name} & \textbf{Language Covered} & \textbf{Unique Speakers} \\ 
    \midrule
    JVS         & Japanese                           & 100   \\ 
    CallHome    & German, Spanish, Chinese, Japanese      & 120   \\ 
    Voxconverse & English                             & 150   \\ 
    \bottomrule
    \end{tabular}
\end{table}
\vspace{-5pt}
During preprocessing, all audio samples were resampled to 16 kHz and processed using a pretrained Whisper feature extractor. Each input is standardized by zero-padding or truncating to 3000 frames, ensuring compatibility with the Whisper encoder.
\subsection{Online Hard Triplet Mining Strategy}
 The training dataset is composed of audio samples with associated speaker labels:
\begin{equation}
\mathcal{D} = \{ (x_i, y_i) \},
\end{equation}
where \(x_i\) denotes an audio sample and \(y_i\) is its corresponding speaker label.

During training, for each audio sample \(x_i\), two augmented views are generated:
 \(x_i^{(n)}\) noise-augmented view,\(x_i^{(t)}\) time-stretched view. Rather than pre-constructing triplets, triplet selection is performed online within each mini-batch. Given a mini-batch \(\mathcal{B} = \{(x_i, y_i)\}_{i=1}^{B}\), each sample acts as an anchor. For each anchor:
\begin{itemize}
    \item A positive sample is selected from the batch as another sample with the same speaker label (\(y_i = y_j\)) that is most dissimilar (i.e., with the maximum Euclidean distance) among available positives.
    \item A negative sample is selected as one with a different speaker label (\(y_i \neq y_k\)) that is most similar (i.e., with the minimum Euclidean distance) among available negatives.
\end{itemize}
This online hard triplet mining strategy generates challenging triplets that drive the learning of discriminative speaker embeddings. Additionally, the self-supervised NT-Xent loss is computed between the original and augmented views, enforcing consistency and further enhancing the robustness of the embeddings.

\subsection{Training}
Our model builds on the \texttt{openai/whisper-tiny} architecture, leveraging its encoder to extract robust audio representations. The encoder output is aggregated via mean pooling and then passed through a projection head composed of two dense layers with ReLU activation, mapping the features to an embedding space of dimension 256. In addition to an online hard triplet loss with a margin of 1.0 to enhance class separation, the training incorporates multi-view self-supervised learning. Two augmented versions of the input audio—one with added Gaussian noise and another with a time-stretch transformation—are generated, and an NT-Xent loss with a temperature of 0.5 is computed to enforce consistency between the original and augmented views. The overall loss is a combination of the triplet loss and the self-supervised NT-Xent loss (with a weight of 1.0 for the latter). Training is performed in mini-batches of 16 samples over 3 epochs using the Adam optimizer with a learning rate of \(1 \times 10^{-5}\). 

\begin{table}[ht]
    \centering
    \captionsetup{font=small}
    \caption{Updated Training Configuration}
    \label{tab:training_config}
    \small
    \begin{tabular}{@{}lc@{}}
        \toprule
        \textbf{Parameter}                       & \textbf{Value} \\
        \midrule
        Audio Sampling Rate                      & 16 kHz \\
        Batch Size                               & 16 \\
        Epochs                                   & 3 \\
        Learning Rate                            & \(1 \times 10^{-5}\) \\
        Optimizer                                & Adam \\
        Embedding Dimension                      & 256 \\
        Triplet Loss Margin                      & 1.0 \\
        Self-Supervised Loss Weight              & 1.0 \\
        NT-Xent Loss Temperature                 & 0.5 \\
        Fixed Input Frames                       & 3000 (zero-padded) \\
        Backbone Architecture                    & \texttt{openai/whisper-tiny} \\
        Data Augmentations                       & Gaussian noise, Time-stretch \\
        \bottomrule
    \end{tabular}
\end{table}
\vspace{-5pt}


\section{Results and Discussion}
\begin{figure}[!h]
    \centering
    \includegraphics[width=\columnwidth]{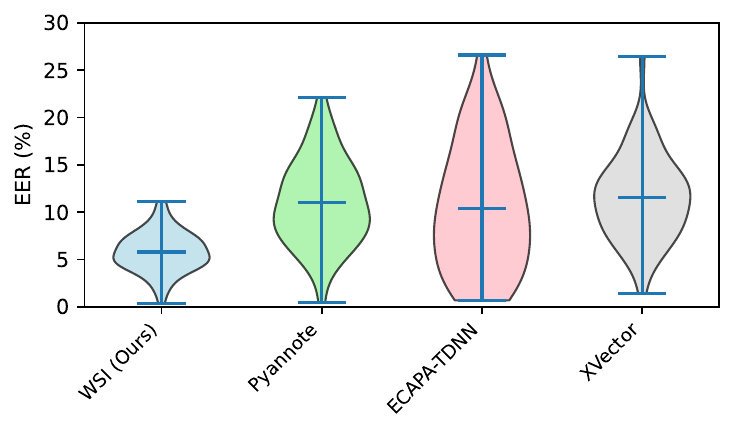}
    \caption{EER Across Methods and Datasets.}
    \label{fig:eer_boxplot}
\end{figure}

In this section, we compare the proposed WSI approach against three baselines (Pyannote Embedding, ECAPA-TDNN, and X-vector) on multiple datasets and languages. Figure~\ref{fig:eer_boxplot} illustrates the distribution of EER values for each method across different different datasets, while Table~\ref{tab:combined_results} provides the detailed numerical results.

On the multilingual VoxTube dataset, WSI achieves an EER of 0.90\% (±0.95), substantially outperforming Pyannote Embedding (3.38\%) and improving upon ECAPA-TDNN (1.17\%) and X-vector (7.23\%). These gains underscore the effectiveness of Whisper’s multilingual pre-trained representations in combination with triplet loss-based fine-tuning.
 \vspace{-5mm}

\begin{table*}[ht]
\centering
\footnotesize  
\caption{Performance Comparison of the Proposed WSI Approach with Various Baselines.}
\label{tab:combined_results}
\setlength{\tabcolsep}{5pt}  
\renewcommand{\arraystretch}{1.1}  
\begin{tabular}{@{}llccccc@{}}
\toprule
\textbf{Dataset} & \textbf{Language} & \textbf{Metric} & \textbf{WSI} & \textbf{Pyannote} & \textbf{ECAPA-TDNN} & \textbf{XVector} \\
\midrule
\multirow{2}{*}{\textbf{VoxTube}} & \multirow{2}{*}{\textit{Multilingual}} 
    & EER ($\downarrow$)      & 0.90 (\,\scriptsize\textit{±0.95}\,)           & 3.38 (\,\scriptsize\textit{±1.64}\,)          & 1.17 (\,\scriptsize\textit{±1.01}\,)           & 7.23 (\,\scriptsize\textit{±4.13}\,)      \\
& & AUC ($\uparrow$)   & 0.99 (\,\scriptsize\textit{±0.009}\,)            & 0.97 (\,\scriptsize\textit{±0.016}\,)           & 0.98 (\,\scriptsize\textit{±0.010}\,)            & 0.92 (\,\scriptsize\textit{±0.041}\,)      \\
\midrule
\multirow{2}{*}{\textbf{JVS}} & \multirow{2}{*}{\textit{Japanese}} 
&  EER ($\downarrow$)        & 8.48 (\,\scriptsize\textit{±3.55}\,) & 26.39 (\,\scriptsize\textit{±7.70}\,) & 21.22 (\,\scriptsize\textit{±3.65}\,) & 21.23 (\,\scriptsize\textit{±3.66}\,) \\
& & AUC ($\uparrow$)   & 0.79 (\,\scriptsize\textit{±0.015}\,) & 0.78 (\,\scriptsize\textit{±0.020}\,) & 0.78 (\,\scriptsize\textit{±0.018}\,) & 0.77 (\,\scriptsize\textit{±0.022}\,) \\
\midrule
\multirow{2}{*}{\textbf{CallHome}} 
& \multirow{2}{*}{\textit{German}} 
&  EER ($\downarrow$)        & 5.50 (\,\scriptsize\textit{±2.50}\,) & 15.30 (\,\scriptsize\textit{±3.20}\,) & 11.00 (\,\scriptsize\textit{±2.70}\,) & 12.00 (\,\scriptsize\textit{±2.80}\,)\\
& & AUC ($\uparrow$)   & 0.93 (\,\scriptsize\textit{±0.020}\,) & 0.88 (\,\scriptsize\textit{±0.025}\,) & 0.91 (\,\scriptsize\textit{±0.022}\,) & 0.90 (\,\scriptsize\textit{±0.023}\,)\\
\cmidrule{2-7}
\multirow{2}{*}{\textbf{CallHome}} 
& \multirow{2}{*}{\textit{Spanish}} 
&  EER ($\downarrow$)        & 9.20 (\,\scriptsize\textit{±2.40}\,) & 16.00 (\,\scriptsize\textit{±3.30}\,) & 11.50 (\,\scriptsize\textit{±2.50}\,) & 12.50 (\,\scriptsize\textit{±3.00}\,)\\
& & AUC ($\uparrow$)   & 0.92 (\,\scriptsize\textit{±0.022}\,) & 0.87 (\,\scriptsize\textit{±0.026}\,) & 0.90 (\,\scriptsize\textit{±0.023}\,) & 0.89 (\,\scriptsize\textit{±0.024}\,)\\
\cmidrule{2-7}
\multirow{2}{*}{\textbf{CallHome}} 
& \multirow{2}{*}{\textit{Chinese}} 
&  EER ($\downarrow$)        & 6.00 (\,\scriptsize\textit{±2.60}\,) & 17.00 (\,\scriptsize\textit{±3.40}\,) & 12.50 (\,\scriptsize\textit{±2.70}\,) & 13.50 (\,\scriptsize\textit{±3.10}\,)\\
& & AUC ($\uparrow$)   & 0.91 (\,\scriptsize\textit{±0.024}\,) & 0.86 (\,\scriptsize\textit{±0.027}\,) & 0.89 (\,\scriptsize\textit{±0.025}\,) & 0.88 (\,\scriptsize\textit{±0.026}\,)\\
\cmidrule{2-7}
\multirow{2}{*}{\textbf{CallHome}} 
& \multirow{2}{*}{\textit{Japanese}} 
&  EER ($\downarrow$)        & 6.80 (\,\scriptsize\textit{±2.40}\,) & 15.00 (\,\scriptsize\textit{±3.10}\,) & 11.00 (\,\scriptsize\textit{±2.50}\,) & 12.20 (\,\scriptsize\textit{±2.90}\,)\\
& & AUC ($\uparrow$)   & 0.93 (\,\scriptsize\textit{±0.021}\,) & 0.88 (\,\scriptsize\textit{±0.024}\,) & 0.91 (\,\scriptsize\textit{±0.022}\,) & 0.90 (\,\scriptsize\textit{±0.023}\,)\\
\midrule
\multirow{2}{*}{\textbf{Voxconverse}} & \multirow{2}{*}{\textit{English}} 
&  EER ($\downarrow$)        & 4.50 (\,\scriptsize\textit{±1.50}\,) & 6.00 (\,\scriptsize\textit{±1.80}\,) & 4.80 (\,\scriptsize\textit{±1.60}\,) & 5.50 (\,\scriptsize\textit{±1.70}\,)\\
& & AUC ($\uparrow$)   & 0.98 (\,\scriptsize\textit{±0.010}\,) & 0.96 (\,\scriptsize\textit{±0.012}\,) & 0.97 (\,\scriptsize\textit{±0.011}\,) & 0.95 (\,\scriptsize\textit{±0.013}\,)\\
\bottomrule
\end{tabular}
\end{table*}

On the monolingual Japanese JVS corpus, WSI records an EER of 8.48\% and an AUC-score of 0.79. Although the absolute error rate is higher due to the complexity of the dataset, WSI still outperforms Pyannote Embedding (26.39\%) ECAPA-TDNN (21.22\%) and X-vector (21.23\%), showing its robustness in language-specific scenarios.

For the CallHome corpus, which includes German, Spanish, Chinese, and Japanese speech, WSI consistently achieves lower EERs and higher AUC-scores than the baselines in each language subset. For instance, in CallHome-German, WSI attains an EER of 5.50\%, outperforming Pyannote Embedding’s 15.30\%, and similar trends are observed in the Spanish, Chinese, and Japanese subsets. Finally, on the English Voxconverse dataset, WSI achieves an EER of 4.50\%, surpassing all competing methods and further confirming its effectiveness under diverse acoustic conditions. Taken together, these results demonstrate that leveraging a multilingual pre-trained ASR encoder with deep metric learning can significantly enhance speaker verification performance in both multilingual and monolingual settings. 
We aslo conduct an ablation study that revealed that the online
hard triplet loss with self-supervised NT-Xent loss plays a crucial role in achieving optimal performance; omitting it increases the Equal Error Rate from 0.90\% to 2.50\% and decreases the AUC score from 0.99 to 0.95.

\section{Conclusion and Future Work}

In this work, we introduced \textit{WSI}, a robust framework that adapts pre-trained acoustic embeddings from the Whisper model for open-set speaker identification. By leveraging Whisper's extensive multilingual pre-training and integrating online
hard triplet loss and a self-supervised loss, WSI achieves exceptional performance across both multilingual and single-language datasets. Our extensive evaluations on the VoxTube, JVS, CallHome, and VoxConverse corpora demonstrate that WSI consistently outperforms established speaker embedding models, attaining lower error rates and higher accuracy in discriminating between speakers. The success of WSI can be attributed to several factors. First, Whisper's pre-training on a diverse set of languages enables the extraction of language-agnostic acoustic features, thereby enhancing the model's generalization across various linguistic contexts. Second, the incorporation of joint loss optimization, resulting in highly discriminative speaker embeddings. 

Despite these promising outcomes, our approach has some limitations. The Whisper encoder is inherently designed to process 30-second audio segments, necessitating zero-padding for shorter inputs. This strategy increases computational overhead and may introduce inefficiencies in real-time applications. Future work will explore alternative strategies, such as modifying the encoder architecture to handle variable-length inputs more effectively without excessive padding.

In summary, WSI represents a significant advancement in speaker identification by effectively combining multilingual pre-trained models with deep metric learning. Its superior performance in both multilingual and single-language settings positions it as a valuable tool for future developments in speaker recognition technology.

\bibliographystyle{IEEEbib}
\bibliography{references} 

\end{document}